\documentclass[a4paper,fleqn,usenatbib]{mnras}

\usepackage{newtxtext,newtxmath}

\usepackage[T1]{fontenc}
\usepackage{ae,aecompl}

\usepackage{graphicx}	
\usepackage{amsmath}	
\usepackage{amssymb}	

\newcommand{\speed}[1]{#1 km~s${}^{-1}$}
\newcommand{\accel}[1]{#1 km~s${}^{-2}$}
\newcommand{\nfig}[1]{Figure~\ref{#1}}

\title[A dispersively formed QFP wave caused by a mini-filament eruption]{Dispersively formed quasi-periodic fast magnetosonic wavefronts due to the eruption of a nearby mini-filament}

\author[Yuandeng Shen]{
Yuandeng Shen$^{1,2,3}$\thanks{E-mail: ydshen@ynao.ac.cn}
Tengfei Song$^{1,4}$
and Yu Liu$^{1,3}$
\\
$^{1}$Yunnan Observatories, Chinese Academy of Sciences,  Kunming, 650216, China.\\
$^{2}$State Key Laboratory of Space Weather, Chinese Academy of Sciences, Beijing 100190, China\\
$^{3}$Center for Astronomical Mega-Science, Chinese Academy of Sciences, Beijing, 100012, China.\\
$^{4}$School of Optoelectronic Engineering, Chongqing University, Chongqing 400044, China
}

\date{Accepted 2018 March 2. Received 2018 March 1; in original form 2017 December 24}

\pubyear{2018}

\begin{document}
\label{firstpage}
\pagerange{\pageref{firstpage}--\pageref{lastpage}}
\maketitle

\begin{abstract}
The observational analysis is performed to study the excitation mechanism and the propagation properties of a quasi-periodic fast-propagating (QFP) magnetosonic wave. The QFP wave was associated with the eruption of a nearby mini-filament and a small B4 {\em GOES} flare, which may indicate that the generation of a QFP wave do not need too much flare energy. The propagation of the QFP wave was along a bundle of funnel-shaped open loops with a speed of about \speed{1100 $\pm$ 78}, and an acceleration of \accel{-2.2 $\pm$ 1.1}. Periodicity analysis indicates that the periods of the QFP wave are $43 \pm 6$, $79 \pm 18$ second. For the first time, we find that the periods of the QFP wave and the accompanying flare are inconsistent, which is different from the findings as reported in previous studies. We propose that the present QFP wave was possibly caused by the mechanism of dispersive evolution of an initially broadband disturbance resulted from the nearby mini-filament eruption.
\end{abstract}




\section{Introduction}
Magnetohydrodynamics (MHD) waves are ubiquitous in the magnetic dominated solar corona. The observations of coronal waves can be used to remote diagnosing physical parameters of the coronal plasma that are hard to measure directly but important for understanding coronal physics, and the energy carried by MHD waves is thought to be a possible source for heating the coronal plasma \citep{nakariakov05a}. The successful detection of coronal waves relies on both spatial and temporal resolution of the observations, namely, the pixel size and time cadence should be smaller and shorter than the wavelength and period of a wave, respectively. However, due to the coarse temporal and spatial resolution observations in the past, besides the global Extreme Ultraviolet (EUV) waves that are believed to be driven by coronal mass ejections  \citep[CMEs; e.g.,][]{shen12a,shen12c,shen13b,shen14a,shen14b,shen17,mei12,xue13,yang13}, people did not find any believable imaging evidence of fast-mode magnetosonic waves before the launch of the {\em Solar Dynamics Observatory} \citep[{\em SDO};][]{pesnell12}. Theretofore, \citet{williams02} reported a fast-mode magnetosonic wave that travels through the loop apex with a velocity of \speed{2100} and a period of 6 second. In addition, \citet{verwichte05} reported a propagating fast magnetosonic kink wave in a vertical open coronal loop with periods in the range of 90 -- 220 second, and the speed ranges from \speed{200 to 700}.

The first detailed unambiguous imaging observation of quasi-periodic fast-propagating (QFP) wave was  reported by \citet{liu11}, by using high resolution images taken by the Atmospheric Imaging Assembly \citep[AIA;][]{lemen12} onboard the {\em SDO}. They found that multiple arc-shaped wave fronts propagate along funnel-shaped coronal loops at a speed of about \speed{2200}. The periods of that QFP wave are 40, 69, and 181 second, in which the last one is consistent with the period of the accompanying flare. Therefore, the authors proposed that the periodicities of the wave and the flare are possibly caused by a common physical regime. Thereafter, the intriguing QFP waves have attracted a lots of attention of solar physicists.  For example, \citet{shen12b} and \citet{shen13a} found that QFP waves not only share common periods with the accompanying flares, but also have some additional periods that cannot be found in the flare. Therefore, the authors proposed that the leakage of the pressure-driven photosphere oscillations to the coronal should be another important mechanism for driving QFP waves. \cite{yuan13} found that the QFP wave presented in \citet{shen12a} was in fact composed by three sub-QFP waves, and each sub-QFP wave was tightly in association with a small radio bursts. This suggests that the excitation of the observed QFP waves are tightly related to the flare energy releasing process \citep{shen18}. In addition, QFP waves are also reported to be associated with CMEs and global EUV waves, which further cause filament oscillations and periodic radio bursts \citep{liu12,shen13a,goddard16}. Other observational studies on QFP waves can be found in articles \citep{nistico14,zhang15,kumar15a,kumar15b,kumar17,qu17}. In addition, \citet{liu14} summarized some basic properties of QFP waves based on the published events before 2014.

Theoretical and numerical studies on QFP waves are also performed in recent years. It has been proposed that QFP waves can be excited by different physical mechanisms. For example, periodic velocity pulsations at the footpoint of the guiding coronal loop \citep{ofman11}, non-linear processes in the magnetic reconnections \citep{kliem00,ofman06,ni12,yang15,takasao16}, and dispersion effect of localized impulsive disturbances \citep{pascoe13,pascoe14,nistico14}. Although these theoretical endeavor, the detailed excitation mechanism of QFP waves is still unclear. Therefore, more observational studies are required for the investigation of the excitation mechanism of QFP waves.

In this paper, a detailed observational analysis of a QFP wave on 2015 July 12 is present to study the  excitation mechanism and the propagation property. For the first time, we find that the periods of the QFP wave and the accompanying flare are different, which is different from the findings that have been reported in previous observational studies, and the mechanism of dispersive evolution of the the disturbance resulted from the nearby mini-filament eruption is proposed as the excitation mechanism of the present QFP wave. It should be noted that in high-resolution observations the eruption of mini-filaments has became more and more important for different kinds of solar eruptions. Many large-scale eruptions are initially caused by the eruption of mini-filaments \citep[e.g.,][]{shen12d,shen17,shen17b,tian17,li17d,li18,hong11,hong17}. The AIA 171 and 304 \AA\ images are used to analyze the event, which has a cadence of 12 s and a pixel size of 0.6 arc-second. The LASCO images are used to show the associated CME, and the soft X-ray fluxes recorded by {\em GOES} are used to study the associated flare. Next section presents the main analyzing results, conclusions and discussions are given in the last section.

\section{Results}
On 2015 July 12, a mini-filament eruption occurred at about 17:34:06 UT close to the northeast limb of the solar disk, which caused a small flare-like bump (B4 level) on the {\em GOES} soft X-ray 1 -- 8 \AA\  flux curve. In addition, flare-like brightenings are observed during the filament eruption. It is interesting that such a miniature filament eruption not only resulted in a fast-mode QFP wave along a bundle of nearby funnel-shaped coronal loop, but also the occurrence of a large-scale CME in the field-of-view (FOV) of the LASCO C2.

\begin{figure}
\includegraphics[width=0.95\columnwidth]{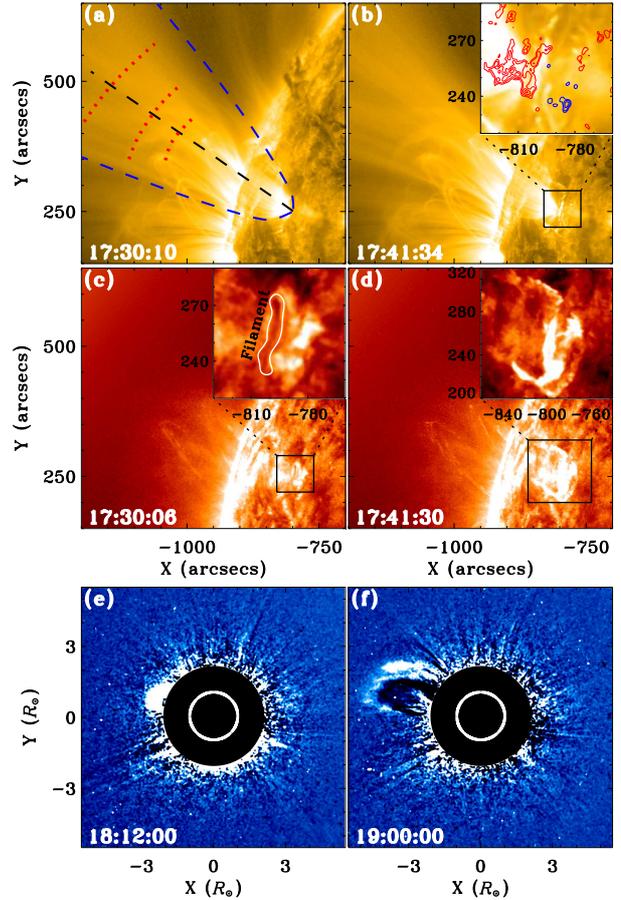}
\caption{The top (middle) row shows the AIA 171 (304) \AA\ images,  while the bottom row shows the LASCO C2 running difference images. The blue dashed curve in panel (a) outlines the funnel-shaped loops, while the dotted red curves mark the positions of the wave fronts at 17:41:22 UT. The black dashed line shows the path along which time-distance diagrams are obtained. The insets in panels (b) -- (d) are close up views of the eruption source region. The mini-filament is outlined in panel (c), and the positive and negative LOS magnetic fields are overlaid in panel (b) as red and blue contours, respectively. In panel (e) and (f), the inner white circle marks the size of the Sun, and the black plate represents the coronagraph's inner occulting disk.}
\label{fig1}
\end{figure}

An overview of the event is shown in \nfig{fig1}. In the AIA 171 \AA\ images, one can see a bundle of funnel-shaped loops which has been outlined by a blue dashed curve as shown in \nfig{fig1} (a). In the AIA 304 \AA\ images, a mini-filament (length $\approx$ 26 Mm) can be observed close to the loops' footpoint, which has a length of about 26 Mm and has been outlined by a white contour in the inset panel in \nfig{fig1} (c). The HMI line-of-sight magnetogram indicates that the filament was located on the magnetic polarity reversion line, and the funnel-shaped loops rooted in a region of positive magnetic field (see \nfig{fig1} (b)). The erupting filament is shown in \nfig{fig1} (d), which erupted to the northeast direction and caused obvious brightenings in the source region. It is interesting that the eruption of the mini-filament further caused a large-scale CME in the FOV of the LASCO C2 (\nfig{fig1} (e) and (f)). According to the measurement of the Coordinated Data Analysis Workshops (CDAW) CME catalog, the first appearance of the CME in the FOV of LASCO C2 was at 18:12:00 UT, and its linear speed and acceleration are \speed{416} and \accel{-7.9}, respectively. 

\begin{figure}
\includegraphics[width=0.95\columnwidth]{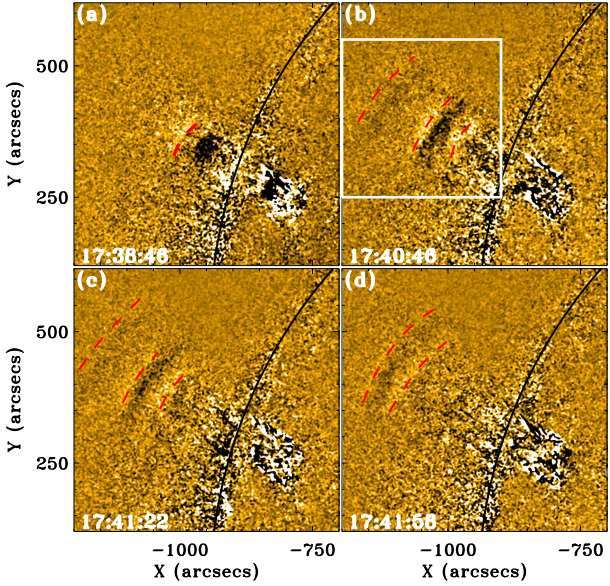}
\caption{AIA 171 \AA\ running difference images show the propagation of the QFP wave. The black curve in each panel indicates the solar limb, while the central position of each wave front is highlighted with a red dashed curve. The white box in panel (b) indicates the region where Fourier analysis is applied. An animation is available in the online journal.}
\label{fig2}
\end{figure}

\nfig{fig2} shows the propagation and evolution of the wave fronts with AIA 171 \AA\ running difference images. Here, a running difference image is obtained by subtracting the present image by the previous one in time, and moving features can be observed clearly in running difference images. The first wave front appeared at 17:38:46 UT at a distance of 158 Mm from the loop's footpoint (\nfig{fig2} (a)). It propagated outward and the following wave fronts successively appeared at the same place. Finally, these wave fronts disappeared at a distance of about 354 Mm from the loop's footpoint. The width of the wave fronts increased rapidly as they propagate outward. We traced the first wave front and found that its width increased from 29 Mm at 17:38:46 UT to 99 Mm at 17:41:22 UT, and the growth rate is about \speed{449}. In addition, it is also measured that the intensity variation relative to the background is about 2\%, which is consistent with the value (1\% -- 5\%) measured in \cite{liu11}.

\begin{figure}
\includegraphics[width=0.95\columnwidth]{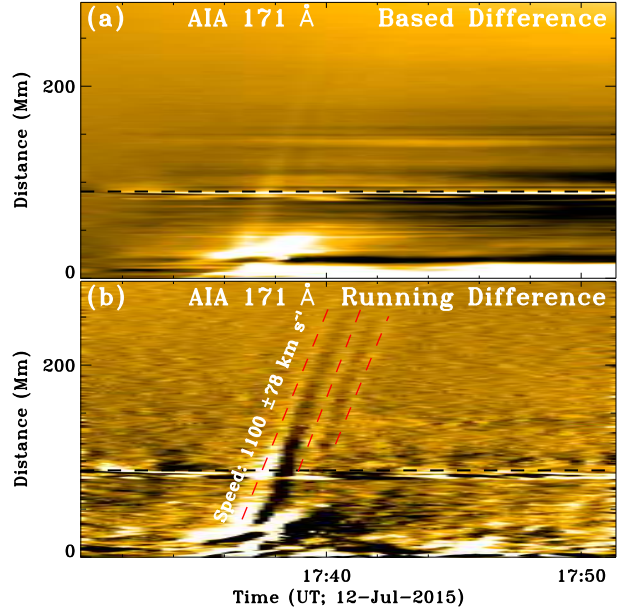}
\caption{Time-distance diagrams show the kinematics of the wave fronts. Panels (a) and (b) are the time-distance diagrams made from AIA 171 \AA\ based and running difference images along the black dashed line as shown in \nfig{fig1} (a), respectively. The horizontal black dashed line in each panel indicates the position of the solar limb, and the red dashed lines in panel (b) are the linear fit to the white ridges that represent the propagating wavefronts.}
\label{fig3}
\end{figure}

The kinematics of the wave train is analyzed with time-distance diagrams along the black dashed line as shown \nfig{fig1} (a), and the results are plotted in \nfig{fig3}. To obtain a time-distance diagram, one need to firstly obtain the one-dimensional intensity profiles along the a path at different times, and then a two-dimensional time-distance diagram can be generated by stacking the obtained one-dimensional intensity profiles in time. It can be seen that three inclined white ridges can be clearly identified in the time-distance made from AIA 171 \AA\ running difference images, but they are very weak in the time-distance made from AIA 171 \AA\ based difference images. This is possibly due to the lower intensity variation relative to the background. The observed white ridges in the time-distance diagrams represent the three observed wave fronts as shown by the AIA 171 images. By fitting each ridge with both linear and second order polynomial functions, it is obtained that the linear speed of the three wave fronts range from \speed{1064 to 1152}, and the acceleration is in the range of \accel{-1.0 -- -3.4}. It is calculated that the average speed and acceleration are of \speed{1100 $\pm$ 78} and \accel{2.2 $\pm$ 1.1}, respectively.

\begin{figure}
\includegraphics[width=0.95\columnwidth]{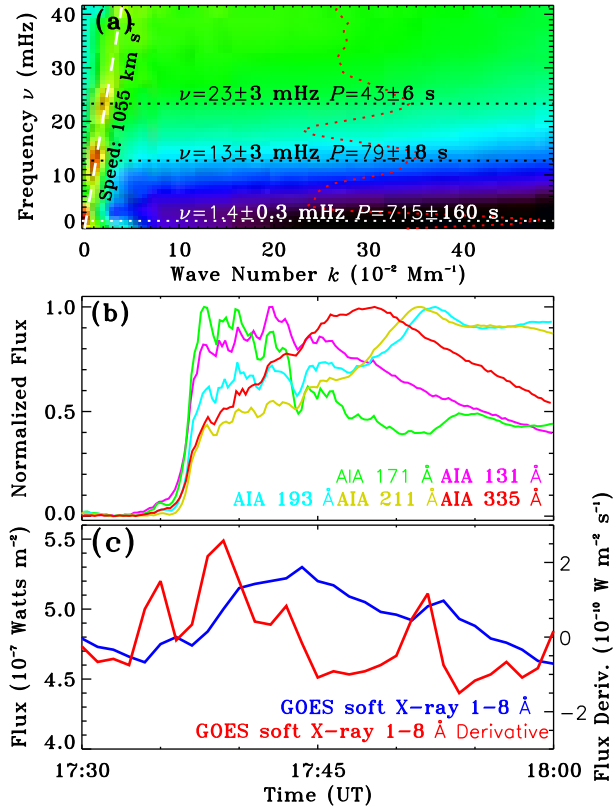}
\caption{Panel (a) shows the Fourier power spectra ($k$--$\omega$ diagram) of a three-dimensional data cube of AIA 171 \AA\ running difference images during 17:30:00 UT -- 17:45:00 UT in the white box region as shown in \nfig{fig2} (b). The horizontal dotted lines indicate the possible frequencies of the QFP wave, which are determined by the peaks showing on the intensity profile of the straight ridge (red curve). The white dashed line is a linear fit to the ridge. Panel (b) shows the lightcurves of the eruption source region measured from the AIA's different channels. Panels (c) shows the {\em GOES} soft X-ray flux in the energy band of 1 -- 8 \AA\ (blue) and its derivative (red), respectively.}
\label{fig4}
\end{figure}

To analyze the periodicity of the QFP wave, the Fourier power spectra ($k$--$\omega$ diagram) \citep{deforest04,liu11,shen12b} is generated using the three-dimensional data cube of AIA 171 \AA\ running difference images (see \nfig{fig4} (a)). A bright ridge passing through the origin of the coordinates can be identified clearly in the $k$--$\omega$ diagram, which represents the dispersion relation of the QFP wave, and its slope represent the wave speed. By applying a linear fit to the ridge, it can be obtained that the speed of the QFP wave is about \speed{1055}, which is in agreement with the value measured from the time-distance diagrams. In addition, there are three dense nodes along the ridge, which represent the possible frequencies (periods) in the QFP wave. It is measured that the frequencies of the wave are $1.4 \pm 0.3$, $13 \pm 3$, and $23 \pm 3$ mHz (see the horizontal dotted lines in \nfig{fig4} (a)), and the corresponding periods are $715 \pm 160$, $79 \pm 18$, and $43 \pm 6$ second. Here, the error for each frequency is determined by the full-width at half-maximum of each node in the intensity curve of the ridge as shown by the red dotted curve in \nfig{fig4} (a). Since the lifetime of the QFP wave is about 5 minutes from 17:37:00 to 17:42:00 UT, the frequency of $1.4 \pm 0.3$ ($715 \pm 160$ s) mHz should be excluded since this period is larger that the lifetime of the wave. Therefore, the QFP wave should have two reliable frequencies of  $13 \pm 3$ ($79 \pm 18$ second), and $23 \pm 3$ ($43 \pm 6$ second) mHz. Here, it is noted that the period of $79 \pm 18$ second is similar to the one observed in \cite{shen13a}, which may reflect the similar property of the plasma medium which support the propagation of QFP waves.

The flare lightcurves obtained from AIA's different channels in the box region as shown in \nfig{fig1} (c) are plotted in \nfig{fig4} (b), while the {\em GOES} 1 -- 8 \AA\ soft X-ray flux and its derivative are plotted in \nfig{fig4} (c). One can see that the {\em GOES} soft X-ray flux only shows a small bump during the eruption of the mini-filament, but the rising of the AIA lightcurves are obvious. In addition, we note some pulsations in the AIA lightcurves and the derivative {\em GOES} soft X-ray flux (see \nfig{fig4} (b) and (c)). We find that the possible periods in the flare are very different with those of the QFP waves. This result is different from the previous findings that QFP waves always share similar periods with the accompanying flares.

\section{Conclusions \& Discussions}
Using high temporal and high spatial resolution imaging observations taken by the {\em SDO}/AIA, we present the detailed analysis of a QFP wave on 2015 July 12, which was associated with a mini-filament eruption and a small B4 {\em GOES} flare. It is interesting that such a less energetic solar eruption also caused a large-scale CME, which is different with the scenario that mini-filament eruptions cause mini-CMEs \citep{hong11}. In addition, this event still indicates that the generation of a QFP wave do not require too much flare energy. The QFP wave showed three arc-shaped wave fronts propagating along a bundle of funnel-shaped coronal loops that rooted close to the mini-filament. The QFP wave started during the eruption phase of the mini-filament, whose appearance and disappearance distances from the loop's footpoint are about 158 and 354 Mm, respectively. During the propagation, the width of the wave front increased rapidly from 29 Mm to 99 Mm with a growth rate of about \speed{449}, and the intensity variation relative to the coronal background is about 2\%. It is measured that the linear speed of the QFP wave is about \speed{1100 $\pm$ 78}, and the acceleration is about \accel{-2.2 $\pm$ 1.1}. Fourier analysis indicates that the QFP wave has two reliable periods of $79 \pm 18$ and $43 \pm 6$ second. In addition, it is found that the possible periods in the flare are very different with those of the QFP waves.

For the first time, it is found that the periods of the QFP wave and the accompanying flare are very different, which indicates that the excitation of the QFP wave and the periodicity of the accompanying flare should be caused by different physical mechanisms. In previous studies it is found that QFP wave and the accompanying flares often share some common periods. Therefore, previous studies often proposed that the generation of QFP waves and the periodicity of the accompanying flares are excited by a common physical mechanism \citep[e.g.,][]{liu11,liu12,shen12b,shen13a}. It is proposed that the excitation mechanism of QFP waves should be tightly related to the nonlinear processes in the magnetic reconnection that produce flares. For example, the periodic coalescence and separation of plasmoids in the reconnection current sheet and their interaction with the ambient magnetic field lines \citep[e.g.,][]{kliem00,ni12,yang15,takasao16}, and the presence of shear flows in the current layer \citep{ofman06}. In addition, \citet{shen12b} found that some periods in the QFP wave are not consistent with the flare, but similar to the photospheric pressure-driven oscillations. Therefore, the authors proposed that the leakage of photospheric oscillation to the coronal is also important for exciting QFP waves.

For the present QFP wave, the different periods of the QFP wave and the accompanying flare suggests that the QFP wave should be excited by another physical mechanism different from what has been proposed in previous studies. Before the imaging observations of QFP waves, quasi-periodic pulsations (QPPs) are frequently detected in solar and stellar flares from radio to hard X-ray waveband \citep[e.g.,][]{tsiklauri04,mitra05,nakariakov09,inglis12,van16,li15,li17a,li17b,li17c}. Therefore, we can draw on the experience of QPPs in flares and find some clues to understand the excitation mechanism of the present QFP wave. \cite{nakariakov05b} summarized four possible mechanisms for driving QPPs, including 1) geometrical resonances, 2) dispersive evolution of initially broadband signals, 3) nonlinear processes in magnetic reconnections, and 4) the leakage of oscillation modes from other layers of the solar atmosphere. In the present event, it is found that the evolution of the QFP wave is similar to the theoretical prediction and numerical experiments that a fast-mode magnetosonic wave can be excited by the dispersive evolution of a localized impulsive disturbance in a straight magnetic structure, which often experiences three distinct phases: periodic, quasi-periodic, and decay phases \citep{roberts83,roberts84,murawski93a,murawski93b,nistico14}. The periodic phase cannot be observed due to the lower amplitude, but the wave trains becomes visible during the quasi-periodic phase since the increased amplitude. In view of these possible mechanisms and the similar evolution stages with the dispersive evolution of a localized impulsive disturbance, we propose that the dispersive evolution of an initially broadband disturbance is probably the excitation mechanism of the present QFP wave, in which the disturbance resulted from the impact of the erupting filament upon the nearby funnel-shaped coronal loops.  

\section*{Acknowledgements}
The author thanks the observations provided by the {\em SDO}, and the referee's valuable comments and suggestions. This work is supported by the Natural Science Foundation of China (11403097,11773068,11633008,11533009), the Yunnan Science Foundation (2015FB191,2017FB006), the Specialized Research Fund for State Key Laboratories, the Youth Innovation Promotion Association (2014047) of Chinese Academy of Sciences Sciences, and the grant associated with the Project of the Group for Innovation of Yunnan Province.

\bsp	
\label{lastpage}
\end{document}